\documentclass[9pt, conference]{IEEEtran}
\IEEEoverridecommandlockouts
\usepackage{cite}
\usepackage{amsmath,amssymb,amsfonts}
\usepackage{algorithmic}
\usepackage{textcomp}

\usepackage{setspace}
\def\smallerspacecaption{\vspace{-2mm}}
\def\smallspaceenum{\vspace{1.25mm}}

\usepackage{floatrow}
\usepackage[caption=false,font=footnotesize]{subfig}

\usepackage[table, dvipsnames]{xcolor}  % Modern color package (used for text and tables)
\usepackage{fancyhdr}
\usepackage{graphicx}
\usepackage{url}
\usepackage{verbatim}
\usepackage{multirow}
\usepackage{hhline}
\usepackage[us]{datetime}
\usepackage{paralist}
\usepackage[implicit=false,hidelinks]{hyperref}
\usepackage{stfloats}
\usepackage[ruled, vlined, norelsize]{algorithm2e} % The algorithms writing package
\SetAlFnt{\sf} % Set the algorithms font style to sans serif

\graphicspath{{incl/}}

\usepackage{soul}
\newcommand{\drop}[1]{\textcolor{blue}{\st{#1}}}
\renewcommand{\drop}[1]{}

\newdimen\arrayruleHwidth
\makeatletter
\def\Hline{\noalign{\ifnum0=`}\fi\hrule \@height \arrayruleHwidth
\futurelet \@tempa\@xhline}
\makeatother

\makeatletter
\def\blfootnote{\xdef\@thefnmark{}\@footnotetext}
\makeatother

\shortdate
\settimeformat{ampmtime}

\makeatletter
\newcommand*\tabsize{%
	   \@setfontsize\tabsize{6}{7.2}%
}
\makeatother

\renewcommand{\arraystretch}{1.03}

\renewcommand{\IEEEbibitemsep}{0.3pt}
\makeatletter
\IEEEtriggercmd{\reset@font\normalfont\fontsize{6.8pt}{7.11pt}\selectfont}
\makeatother
\IEEEtriggeratref{1}

\begin{document}

\title{Obfuscating the Interconnects: Low-Cost and Resilient Full-Chip Layout Camouflaging\\
}
%% ACM template

%\numberofauthors{6} 

%  in this sample file, there are a *total*
\newcommand*\samethanks[1][\value{footnote}]{\footnotemark[#1]}

\author{
	{\Large Satwik Patnaik$^{\dagger}$, Mohammed Ashraf\,$^{\ddagger}$, Johann Knechtel$^{\ddagger}$, and Ozgur Sinanoglu$^{\ddagger}$}\\[2pt]
  {\large $^{\dagger}$\,Tandon School of Engineering, New York University, New York, USA}\\
  {\large $^{\ddagger}$\,New York University Abu Dhabi, Abu Dhabi, United Arab Emirates}\\
  \normalsize{\{sp4012, ma199, johann, ozgursin\}@nyu.edu}
  %\texttt{New York University, NY, USA}
}

%\and
%\and
%\and
%\and
%\and

\maketitle

\renewcommand{\headrulewidth}{0.0pt}
\thispagestyle{fancy}
\pagestyle{fancy}
\chead{
\copyright~IEEE, 2017. This is the author's version of the work. It is posted here for your personal use.
       Not for redistribution. The definitive version of Record was published in
       Proc. International Conference On Computer Aided Design (ICCAD) 2017\\
http://dx.doi.org/10.1109/ICCAD.2017.8203758
}

\begin{abstract}

Layout camouflaging (LC) is a promising
technique to protect chip design intellectual property (IP) from reverse engineers.
Most prior art,
however, cannot leverage
the full potential of LC due to excessive overheads and/or their limited scope on an FEOL-centric and accordingly customized manufacturing process.
If at all, most existing techniques can be reasonably applied only to selected parts of a chip---we argue that such ``small-scale or custom
camouflaging'' will eventually be circumvented,
irrespective of the underlying technique.

In this work, we propose a novel LC scheme which is low-cost and generic---full-chip LC can finally be realized without any reservation.
Our scheme is based on obfuscating the interconnects (BEOL); it can be readily applied to any design without modifications in the device layer (FEOL).
Applied with split manufacturing in conjunction, our approach is the first in the literature to cope with both the FEOL fab and the end-user being untrustworthy.
We implement and evaluate our primitives at the (DRC-clean) layout level; our scheme incurs significantly lower cost than most of the previous
works.
When comparing fully camouflaged to original layouts (i.e., for 100\% LC), we observe on average power, performance, and area overheads of 12\%, 30\%, and 48\%, respectively.

Here we also show
empirically that most existing LC techniques (as well as ours)
can only provide proper resilience against powerful SAT attacks once at least 50\% of the layout is camouflaged---only large-scale LC is practically secure.
As indicated, our approach can deliver even 100\% LC at acceptable cost.
Finally, we also make our flow publicly available, enabling the community
to protect their sensitive designs.

\end{abstract}

\section{Introduction}
\label{sec:introduction}

Ensuring the security and trustworthiness of hardware has become a major concern in recent years~\cite{mccants11, skorobogatov12_chapter, rostami14}.
One reason for protecting the hardware is that intellectual property (IP) can otherwise be duplicated without consent, resulting in a financial loss for the IP owner.
Furthermore, understanding the gate-level implementation of a chip may advance other attacks such as side-channel analysis or Trojan insertion~\cite{skorobogatov12_chapter, rostami14}.
A malicious end-user, i.e., an adversary without direct access to the design and fabrication process,
has to resort to \emph{reverse engineering (RE)} of chips to obtain the IP.
The tools and know-how for RE attacks are becoming more advanced and widely available, thus rendering RE a practical
threat~\cite{
	rostami14,
	skorobogatov12_chapter,subramanyan14,sugawara14,courbon16}.

The goal of \emph{layout camouflaging (LC)} is to mitigate RE attacks.\footnote{\emph{Hardware/layout obfuscation} are wide-spread synonyms for LC. 
We use the term \emph{obfuscation} purposefully in the context of \emph{obfuscating the interconnects}, which is the key principle of our work.
Besides, LC is closely related to the concept of \emph{logic locking}~\cite{roy10},
which itself is also known as \emph{logic encryption}.
}
The fundamental idea
is to alter the behavior or appearance of a chip such that it is arduous or even impossible for the RE attacker to infer the chip's real functionality.
This can be achieved, e.g., by ``look-alike'' or ambiguous gates~\cite{cocchi14, rajendran13_camouflage}, by
secretly configured multiplexers (MUXes)~\cite{wang16_MUX,zhang16},
or by threshold-dependent camouflaging of gates~\cite{collantes16,nirmala16,erbagci16}.
We discuss the prior art further in Sec.~\ref{sec:background};
besides,
a comprehensive overview on LC
is given in~\cite{vijayakumar16}.

\textbf{On the cost of prior art:}
Most existing LC schemes have a high layout cost and are accordingly limited for practical use.  For example, the ambiguous XOR-NAND-NOR gate proposed
in~\cite{rajendran13_camouflage} has 5.5$\times$ power, 1.6$\times$ delay, and 4$\times$ area cost in comparison to a conventional NAND gate.
Even promising works such as the threshold-dependent, full-chip LC proposed in~\cite{erbagci16} induces overheads of 14\%, 82\%, and 150\% in power, performance, and area
	(PPA),
     respectively.
Besides, FEOL-based LC techniques are typically only available as customized IP,
and/or require some alterations for the FEOL
manufacturing process,
incurring financial cost on top of PPA overheads.

In practice, {
	existing LC schemes can be applied only selectively---if at all---due to their inherent PPA overheads
and their impact on the FEOL processing.}
As a result, the constrained application of these techniques may lead to a compromise in their security.

\textbf{On the resilience of prior art:}
When LC techniques are applicable only to parts of a chip, the challenge is where and to what
	extent camouflaging shall be effected.
Ideally,
	an attacker's effort to obtain the design from a carefully camouflaged netlist would be exponential in the number of camouflaged
gates~\cite{massad15}.
Advances on the \emph{Boolean satisfiability problem (SAT)}, however, have enabled 
powerful attacks on LC (and on logic locking)~\cite{subramanyan15,massad15,liu16}.
The previously unforeseen success of such SAT attacks stems from the typically small number of input/output (I/O) patterns required in practice for de-camouflaging.
Recent works~\cite{li16_camouflaging, yasin16_CamoPerturb, xie16_SAT} aim for exponentially scaling and \emph{provably secure LC} but are still prone to other advanced attacks (Sec.~\ref{sec:background}).

Accounting for the recent
	advances of analytical and invasive attacks (Sec.~\ref{sec:background}),
we make the following case: {
		to remain resilient, at least as long as foreseeable, LC has to be applied at a large scale}, i.e., more than 50\% of the layout should be camouflaged.

\begin{figure}[tb]
\centering
\includegraphics[height=5.50em]{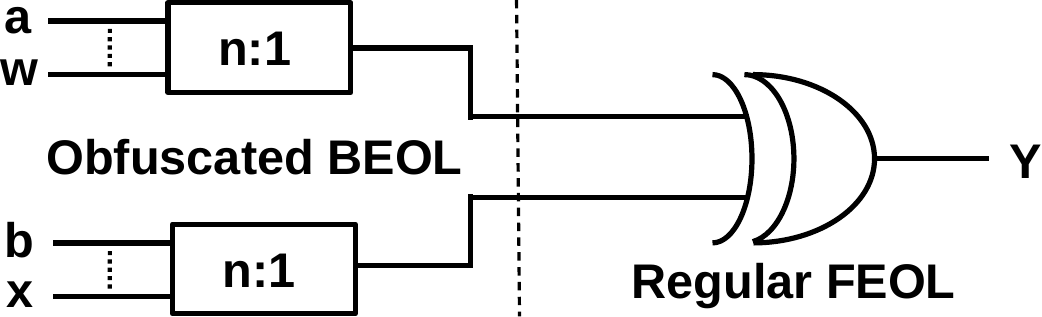}
\smallerspacecaption
\caption{Our camouflaging concept is based on secret $n$:1 mappings in the BEOL, which obfuscate the inputs for any regular gate (not only two-input XOR).
		The set of possible functionalities
		depends on $n$:1, the selection of the input nets, and the gate type.
}\label{fig:concept}
\end{figure}

\textbf{On this paper:}
Here we
promote full-chip LC with high resilience and low cost.
To do so, we propose and evaluate
	simple but effective LC primitives which are based on obfuscating (parts of) the interconnects.
Our contributions
can be summarized as follows:
\smallspaceenum
\begin{compactenum}
\item We enable resilient and full-chip LC, based on a novel, security- and cost-driven approach for \emph{obfuscating the interconnects}.
\item Our novel BEOL-centric LC primitives are tailored for regular gates (Fig.~\ref{fig:concept}).
That is, our primitives do not require any modification at the FEOL device layer and can, therefore, be easily integrated into any (industrial) design
flow. We
make our flow (based on \emph{Cadence Innovus}) publicly available in~\cite{webinterface}.
\item The fact that our primitives implement BEOL-centric LC suggests the (optional) application of \emph{split manufacturing}~\cite{mccants11, rostami14} in conjunction.
Doing so allows us for the first time to hinder fab adversaries in addition to malicious end-users, all while imposing only 
low commercial cost, which may be even compensated for.
Note that the implementation effort and cost of protecting against a malicious fab \emph{and} malicious end-users have been considered as mutually exclusive so far.
\item We assess the resilience of different flavors of our proposed primitives and compare them against previous works on LC. In that process,
we employ
powerful SAT attacks on traditional benchmarks (which are relatively small) while we also
show---for the first time---attacks on
	large VLSI benchmarks.
Besides,
	we introduce the
	notion of
\emph{practically secure LC}, which seeks to impose an excessive computational cost on SAT-based attacks without inserting additional, dedicated circuit structures.
\item We conduct a thorough evaluation of camouflaged, DRC-clean  layouts.
In contrast, most of the previous works
investigate their LC primitives only as stand-alone devices, without applying them in actual layouts; we argue that this is overly optimistic.
Our work is one of the very few in the literature providing comprehensive layout-level evaluation in general, and the first to do so for large VLSI benchmarks
with up to 39,014 gates.
\end{compactenum}

\section{Background}
\label{sec:background}

Next, we discuss the recent progress on LC (along with demonstrated and potential attacks), which is typically focused on FEOL-centric camouflaging. Further, we discuss an early study on
BEOL-centric camouflaging, which also inspired our work to some degree.

\subsection{Camouflaging at the FEOL and Vulnerabilities}
\label{Camouflaging at the FEOL}

As already mentioned, powerful SAT attacks have challenged most prior art on LC (and logic locking)~\cite{subramanyan15,massad15,liu16}.
Thus, several recent works on provably secure LC (and logic locking)~\cite{li16_camouflaging, yasin16_CamoPerturb, xie16_SAT} seek to mitigate
SAT attacks
by inserting 
dedicated (but high-cost) structures which in theory necessitate to consider an
exponential number of I/O patterns.
However, we argue that the tailored structures of~\cite{li16_camouflaging, yasin16_CamoPerturb, xie16_SAT} can be easy to identify during
RE;
 these structures ($i$) are typically applied only in a few places, due to their relatively high cost, and ($ii$)
rely on
arrays of possibly camouflaged gates with their outputs converging in large combinatorial trees.
	Moreover, the output wires of these trees 
	have been successfully identified by signal probability analysis~\cite{yasin17_TETC}.
These critical wires may then be cut to circumvent the security features.\footnote{
Wire cutting has been successfully demonstrated in the past, for example by Helfmeier \emph{et al.}~\cite{helfmeier13}, enabling such invasive attacks in principal.
}
	Besides, advanced SAT attacks have also been demonstrated to mitigate
	provably secure or ``cyclic'' logic locking techniques very recently~\cite{shamsi17,shen17,bypass-attack2017,CycSAT-ICCAD2017,yasin17_TETC}.
In short, a sophisticated attacker may learn which gates to ignore, replace or cut off while (nearly) recovering the original netlist of
such SAT-hardened chips.

RE measures may render FEOL-centric LC also directly void, without the assistance of analytical techniques.
LC schemes such as ``look-alike'' and ambiguous gates~\cite{cocchi14, rajendran13_camouflage} or
secretly configured MUXes~\cite{wang16_MUX,zhang16} rely on dummy contacts or dummy channels.
While it is often claimed by the authors of these studies that simple etching cannot reveal these features, other powerful techniques/tools are available. Specifically, the use of
\emph{scanning electron microscopy} in the \emph{passive voltage contrast} mode (\emph{SEM PVC} in short) allows for accurate and efficient measurement of charge
accumulations; this has been recently demonstrated by Courbon \emph{et al.}~\cite{courbon16} in an unprecedented case study for reading out secured Flash memories.
Now, SEM PVC may break the above LC techniques as well, since dummy contacts/channels will accumulate charges to a much lower degree than real contacts/channels.
Threshold-dependent camouflaging of gates~\cite{collantes16,nirmala16,erbagci16} can also be revealed by SEM PVC,
	as successfully demonstrated
by Sugawara \emph{et al.}~\cite{sugawara14}.
Besides, as the authors in~\cite{collantes16} indicate themselves, monitoring the etch rates can reveal different doping levels which are at the heart of
threshold-dependent gates.

\subsection{Towards Camouflaging at the BEOL}

	Chen \emph{et al.}~\cite{chen15}
suggest implementing vias
either as real, conductive vias (using magnesium, Mg) or as dummy, non-conductive vias (using magnesium oxide, MgO).
That is, the authors advocate  BEOL-centric camouflaging
besides the ``classical LC'' at the FEOL.
Despite their pioneering work on  RE-resilient vias,
Chen \emph{et al.\ }did
not succeed to propose a resilient LC application, as we show in Sec.~\ref{sec:experiments}.
Also, note that our  concept is different from~\cite{chen15}; we only leverage their notion of Mg/MgO vias for obfuscation.

Chen \emph{et al.}~\cite{chen15}
elaborated that the use of Mg/MgO
is practical from both the perspectives of ($i$)~manufacturability and ($ii$)~RE mitigation. For~($i$),
it is noted
that Mg has been traditionally used to facilitate the bonding of copper interconnects to the dielectric layer.
For~($ii$), the authors fabricated samples and observed that Mg was completely oxidized (into MgO)
within a few minutes.
That is, the real Mg vias became indistinguishable from the dummy MgO vias during RE.
Independently, Swerts \emph{et al.}~\cite{swerts15_Mg_BEOL} and Hwang \emph{et al.}~\cite{hwang12_transient_electronics} have used Mg and MgO during CMOS-centric BEOL
processing (without LC in mind). Hwang \emph{et al.\ }have shown that Mg not only oxidizes but also dissolves quickly---as does MgO---when surrounded by fluids, which is inevitable
in classical RE etching procedures.

Although
one can argue that RE of Mg/MgO vias is somehow possible nevertheless,
such an attack is yet to be demonstrated.\footnote{
For example, we do \emph{not} expect the powerful SEM PVC attack~\cite{courbon16} to be successful here. Once an attacker seeks to measure charges in the individual BEOL layers, she/he will inevitably
	rip up all the interconnects during the
 layer-wise RE process. Hence, a localized lack of charges will hint on any non-functional wire, be it an
	 obfuscated dummy or a ripped-up regular wire.}
In any case, our work relies only on the generic concept of {
		RE-resilient interconnects}, and not on particular materials currently available for high-volume manufacturing. Future interconnects, e.g., based on
carbon/graphene or spintronics~\cite{CNT_TSDM17, naeemi14}, may hinder RE as well.

It is important to
note that RE-resilient interconnects
are implemented not only at the designer's choice but also at the manufacturer's discretion.
That is, this concept requires either split manufacturing~\cite{mccants11, rostami14
} or a trusted BEOL process  in conjunction.

Finally, another aspect of this concept is that its commercial cost may be easily compensated for, even when
	split
manufacturing is  applied.
That is because
it
only requires additional BEOL masks which are, e.g., for M5/M6 3.5--4$\times$ cheaper than the Poly masks at 16nm, according to industry experts.
In contrast, most prior FEOL-centric LC (but not threshold-dependent LC) incur a relatively high cost for the different FEOL masks required.
In general, any FEOL-centric approach demands some alterations of IP libraries and/or for the manufacturing process---such alterations are likely more costly than
	camouflaging at the BEOL.

\section{Our Concept and Threat Model}
\label{sec:concept}
	Our key idea
is the following:
we leverage the concept of RE-resilient interconnects
to design novel, BEOL-centric LC primitives which are applicable to any type of regular gates.

\textbf{Our
		concept:}
		We implement $n$ wires
for each input of the gates to be camouflaged, and each wire has---without loss of generality---some of its vias obfuscated.
In other words,
    we obfuscate the real driving wires of any gate via \emph{secret $n$:1 mappings in the BEOL} (Figs.~\ref{fig:concept}--\ref{fig:final_primitive}).
    As a result, the actual function which a gate implements is obfuscated in a simple yet effective manner;
    the set of possible functionalities
    depends on the selection of the wires, on $n$, and on the gate itself.
Note that our concept is generic and directly applicable for any multi-input gates, and not only for two-input gates.

Since our concept employs obfuscation in the BEOL layers, it can be readily applied in conjunction with split manufacturing, even
	with low commercial cost on top (i.e., at least when splitting at
		higher metal layers---which we do by splitting at M5, see also Sec.~\ref{sec:experiments}).

\textbf{Our threat model:} We assume both the end-user and the fab to be untrusted. The latter is in strong contrast to most prior art on LC that \emph{has to} trust the fab because of their
FEOL-centric techniques.

To hinder fab adversaries, i.e., in particular to protect the secret mappings in the BEOL layers, we leverage split manufacturing.\footnote{In a weaker threat model where the fab is trusted,
	our technique can still be directly applied, just without split manufacturing.}
The goal of malicious end-users is to RE the chip's gate-level layout and identify its secret mappings in the BEOL layers---the latter is challenging and yet to be demonstrated
(Sec.~\ref{sec:background}). Ultimately, both adversaries want to reconstruct the original netlist and its IP.
Towards this end,
end-users can use another working chip copy as an oracle for SAT attacks, whereas fab workers can launch \emph{proximity attacks}~\cite{wang16_sm}.

\section{On Different Flavors of Our Primitive}
\label{sec:flavors}

Here, we shed light on several flavors of our LC scheme.
We assess all flavors by ($i$) their impact on PPA and ($ii$) their
SAT attack resilience.
To do so, we ($i$) conduct a thorough GDSII-level analysis and ($ii$) employ
	powerful SAT attacks as proposed in~\cite{subramanyan15}.
	For the latter, the authors made their attack tool publicly available~\cite{code_pramod}; other recent attacks proposed
	in, e.g.,~\cite{liu16,shamsi17,shen17,bypass-attack2017,CycSAT-ICCAD2017} have not been made available to us.
We camouflage the layouts in the range of 10\% to 100\%, in steps of 10\%. 
   Further setup details are given in Sec.~\ref{sec:experiments}.

Without loss of generality, we exemplarily discuss our primitives when applied for two-input gates in the following.

\subsection{Basic Flavor, with Simple $n$:1 Mappings}

\textbf{2:1 mapping:} We first explore the most basic flavor with two wires for each gate's input, i.e., one dummy wire and one real wire.

This flavor comprises only four functionalities.
For example, with $a,w$ being the wires for one input, and $b,x$ being the wires for the other input (see also Fig.~\ref{fig:concept}), the gate can implement
either $f(a,b)$, $f(a,x)$, $f(w,b)$, or $f(w,x)$, where $f$ is the functionality of the gate.
It is easy to see that even less than four functions are realized when
some wires are driven by the same net. We avoid this---for all flavors---by selecting unique nets for each gate's wires (see
		Sec.~\ref{sec:methodology}).

We expect and observe this flavor to break relatively easily against the attacks
we leverage from~\cite{subramanyan15}.
	For benchmarks
\emph{apex4} and \emph{des}, e.g., the attack terminates within 100--300 seconds, even for large-scale LC (i.e., for 50\%).
Compared to our other flavors, this one has the weakest resilience to SAT attacks.
Hence,
    we did not consider the further application or layout-level evaluation of this basic primitive.

\begin{figure}[tb]
\centering
\includegraphics[scale=.419]{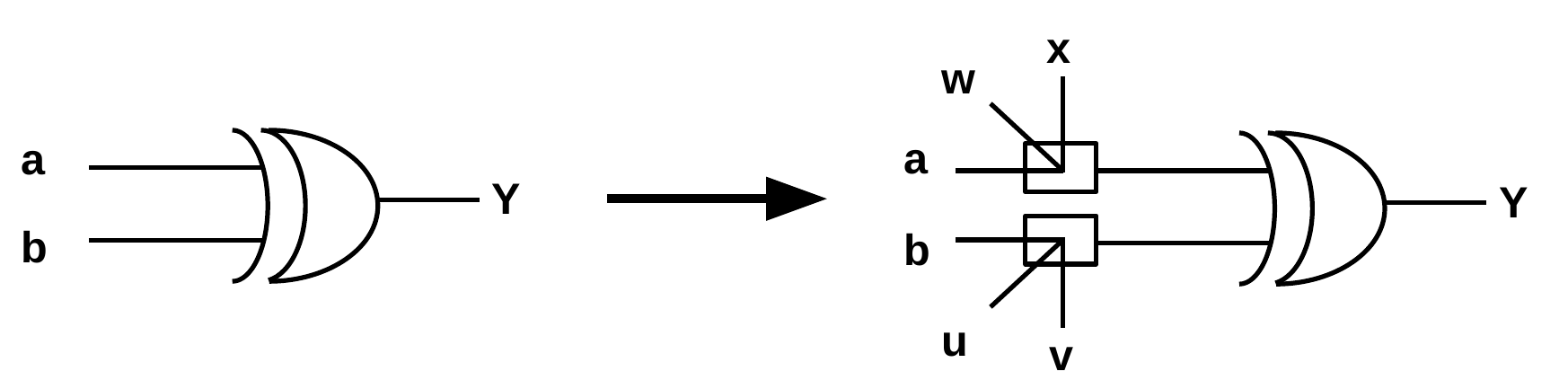}
\smallerspacecaption
\caption{Our basic primitive with secret 3:1 mappings, resulting in up to 9 possible functions for the gate.}\label{fig:basic_primitive}
\end{figure}

\textbf{3:1 mapping:} We extend the basic scheme by adding one more dummy wire for each input (Fig.~\ref{fig:basic_primitive}).
   This primitive
   cloaks one out of
	nine possible functionalities;
	hence, we can expect
   a higher  resilience when compared to the 2:1 mapping primitive.
In fact, we observe that the resilience scales well across all ranges of LC
for this 3:1 primitive and, thus, is enforcing reasonably high efforts for SAT attacks, especially for larger benchmarks such as \emph{b15}.
On average, the attack runtime scales by $\approx$3$\times$ when compared to the 2:1 primitive.

On the flip side, we observe relatively high PPA cost for this primitive, especially for performance/delay; on average,
the delay overhead approaches 60\% when ``only'' 50\% of the layout is camouflaged.
That is because
we need
to choose all the wires
such that they are unique for any gate (as already indicated) and,
as a result,
the more the camouflaging, the higher the routing congestion.
In turn, this increases
wirelength and capacitive loads, which simultaneously aggravates delay and power,
imposing practical limitations for large-scale LC using this basic 3:1 mapping primitive.

\subsection{Extended Flavor, with Fixed Logic Values}
\label{sec:extended_flavor}

\begin{figure}[tb]
\centering
\includegraphics[scale=.416]{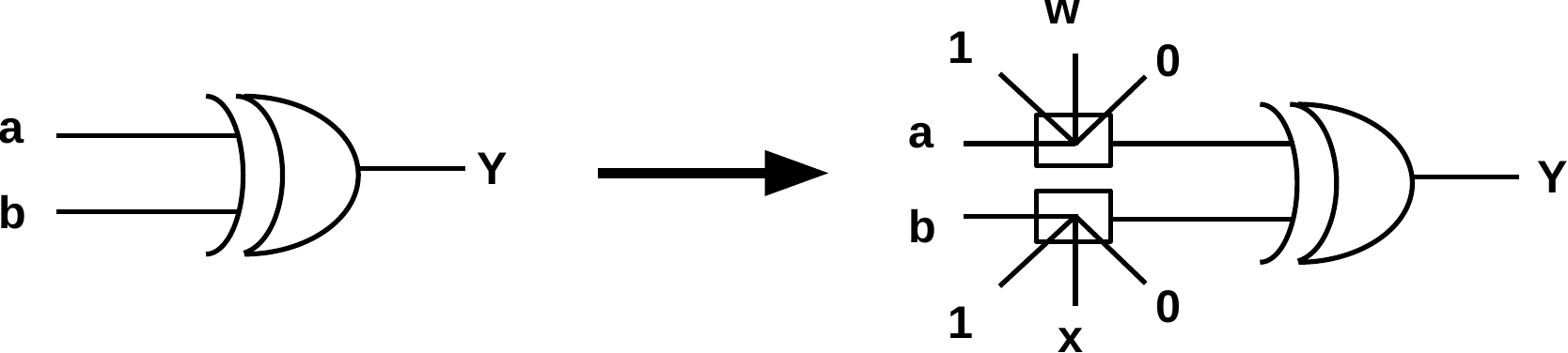}
\smallerspacecaption
\caption{Our extended, final primitive, where the fixed logic values of \emph{0} and \emph{1}
	are employed along with two regular signal nets/wires. Depending on the gate type, 10 or 14 functions are possible.
}\label{fig:final_primitive}
\end{figure}

A promising option
is to additionally employ the fixed logic values \emph{0} and \emph{1} along with regular nets/wires
(Fig.~\ref{fig:final_primitive}).
According to Boolean algebra and considering the underlying gate, this extended primitive provides either $5m$ or $7m$ functionalities, where $m$ is the number of regular wires for each input (in addition to
		the two wires with the fixed values \emph{0} and \emph{1}).
	  For example for $m=2$,
	  with $a,w$ and $b,x$ as the respective regular wires
for the two inputs, and
with XOR as the underlying gate (Fig.~\ref{fig:final_primitive}),
we obtain the following 14 functionalities:
$\emph{0},\emph{1},a,w,b,x,\overline{a},\overline{w},\overline{b},\overline{x}$,
	$a \oplus b$,
	$a \oplus x$,
	$w \oplus b$, and
	$w \oplus x$.

We note that
this approach enables significantly lower PPA cost.
That is, for $m=1$, the average delay overhead is only $\approx$10\%
when 50\% of the layout is camouflaged,
and for $m=2$, the average delay overhead is still only $\approx$20--30\%, even
when 50--100\% of the layout is camouflaged.
When compared to employing only regular nets/wires, the additional use of \emph{0} and \emph{1} offers a fundamental benefit:
their fixed-value wires
are not switching, thus
exhibiting only negligible power consumption and imposing no timing overhead.

We expect the SAT attack resilience of this extended primitive to be
``in between'' the two basic primitives, i.e., at least for $m=1$.
Our experiments corroborate this expectation; for $m=1$, this primitive offers on average a lower resilience than 3:1 mapping (effecting 25--30\% less attack runtime), but a higher
resilience than 2:1 mapping
($\approx$35\% more runtime) when the same set of gates are camouflaged.

To further strengthen the resilience, we may add more regular wires (i.e., increase $m$), thereby extending the set of possible functionalities.
This will directly impact the overall search space and, as a result, increase the average effort required for SAT attacks~\cite{li16_camouflaging}.
Adding more wires (to be driven by unique nets), however, notably contributes to routing congestion which aggravates PPA cost in turn.
In short, we empirically choose $m=2$ for our final primitive (Fig.~\ref{fig:final_primitive}).
We elaborate on this final primitive in Sec.~\ref{sec:experiments} in detail.

\section{Our Camouflaging Methodology}
\label{sec:methodology}

\subsection{Protecting Fixed Values and ``Implausible Functions''}

As indicated above, our final primitive 
appears
attractive
due to its relatively low PPA cost and its adequate resilience (see also Sec.~\ref{sec:experiments}).

For large-scale LC, however, the ubiquitous wires relating to fixed values
may give away clues to an attacker; 
fixed values are typically used only for special registers or ``hardware mode flags.'' An attacker observing a vast number of fixed-value wires---which itself may be easy, given
	that
such wires are typically connected to distinct \emph{TIE cells}---might rightfully
assume that these wires have been introduced for obfuscation.
Thus, unless we
make
these wires
essential for large parts of the design, an
attacker may readily and safely disregard them.

We have to address another, complementary challenge at once.
That is, a mindful attacker may also try to rule out
all
the ``implausible'' functions (i.e., INV, BUF, \emph{0} and \emph{1})
which are those beyond any gate's original functionality.
Since these additional functions arise only due to the fixed values being part of the obfuscated inputs to begin with, they
can only become effective once the fixed values are an essential part of the design (and vice versa).

In short, the fixed-value wires have to be
		rendered essential, while also protecting all the ``implausible functions''
at the same time.\footnote{Independent from our work, Keshavarz \emph{et al.}~\cite{keshavarz17} recently called for maintaining the plausibility of all (chip-level) viable
	functionalities, albeit with a focus on logic synthesis and technology mapping, and without assuming that a working chip is available as an oracle for the (SAT-centric)
		attacks.}
To do so, we perform simple
	\emph{netlist transformations} as follows:
\smallspaceenum
\begin{compactenum}

\item We transform some inverters (INVs) and buffers (BUFs) into
gates of other types (e.g., see Fig.~\ref{fig:primitive_inv}).
Nowadays, around 50\% of all gates are repeaters (INV/BUF)~\cite{shelar13,markov14}, offering ample opportunities for large-scale
transformations/camouflaging.
Here one can freely choose the number of INVs/BUFs to camouflage, and the type of gate (AND, NAND, OR, NOR, XOR, XNOR) to transform them into.
The best strategy, which we also apply, is to randomly transform
50\% of INVs/BUFs.
This way,
an attacker cannot easily infer a direct correspondence between any of those gates and their functionality.
Note how this transformation renders the fixed values essential---they cannot be ignored without misinterpreting the transformed gates. Also, the functionalities INV and
	BUF remain plausible now throughout the layout.

\begin{figure}[tb]
\centering
\includegraphics[scale=.432]{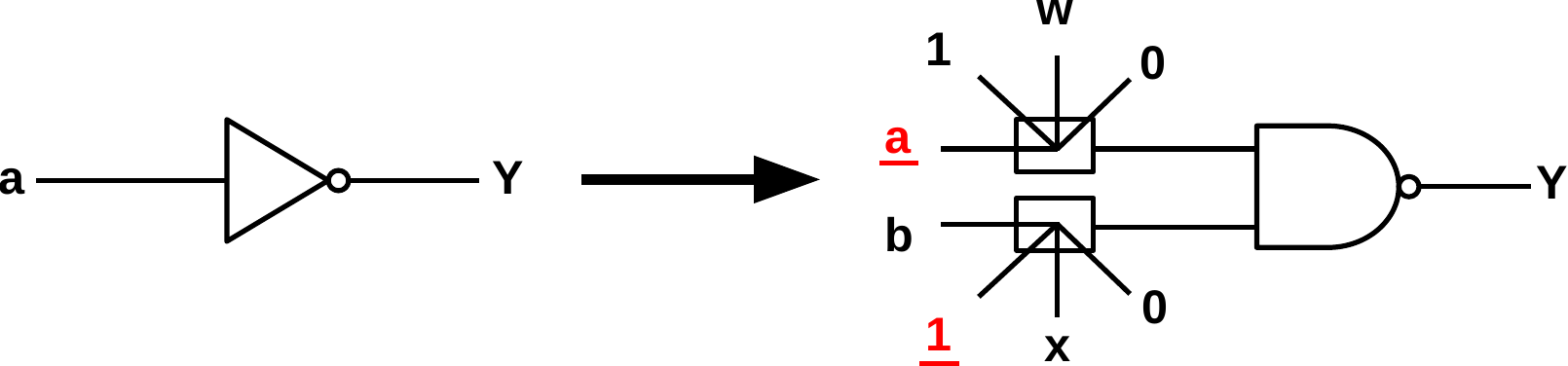}
\smallerspacecaption
\caption{An inverter transformed into a two-input NAND gate.
		The real nets/wires are underlined and shown in red.
}\label{fig:primitive_inv}
\end{figure}

\item We insert some additional gates (into randomly selected regions of whitespace) with
their real inputs tied to fixed values.
These gates act as TIE cells in disguise;
they ``drive'' other camouflaged gates in turn.
Thus, also the functionalities \emph{0} and \emph{1} cannot be ignored anymore without misinterpreting the transformed netlist.

\end{compactenum}

\subsection{Overall Flow}

\begin{figure}[tb]
\centering
\includegraphics[width=.725\textwidth]{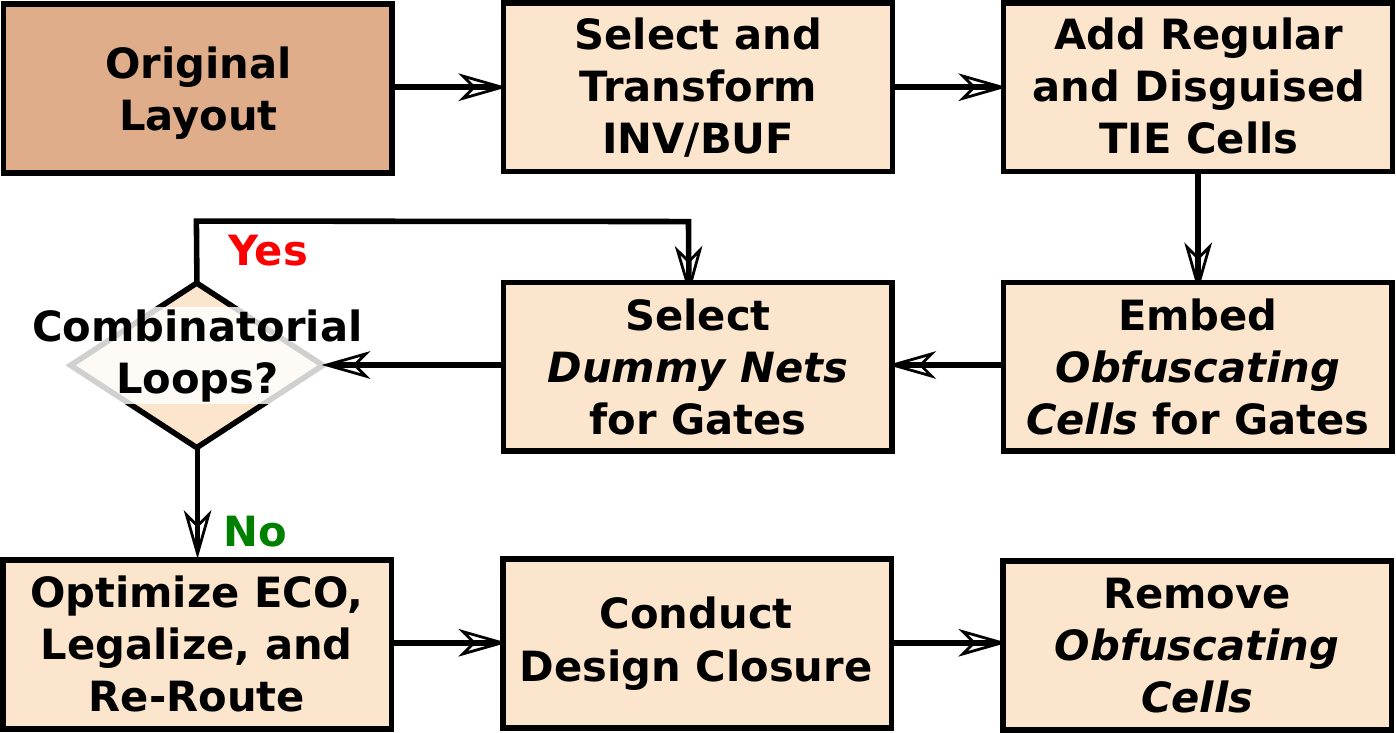}
\smallerspacecaption
\caption{Flow of our layout-level, BEOL-centric camouflaging methodology.}
	\label{fig:flow}
\end{figure}

Here we provide an overview of our camouflaging methodology (Fig.~\ref{fig:flow}), which can be easily
integrated into any design flow.
In this work, we implement our methodology for \emph{Cadence Innovus.} We also provide open access to our flow in~\cite{webinterface}.

\begin{figure}[tb]
\centering
\subfloat[]{\includegraphics[height=2.5cm]{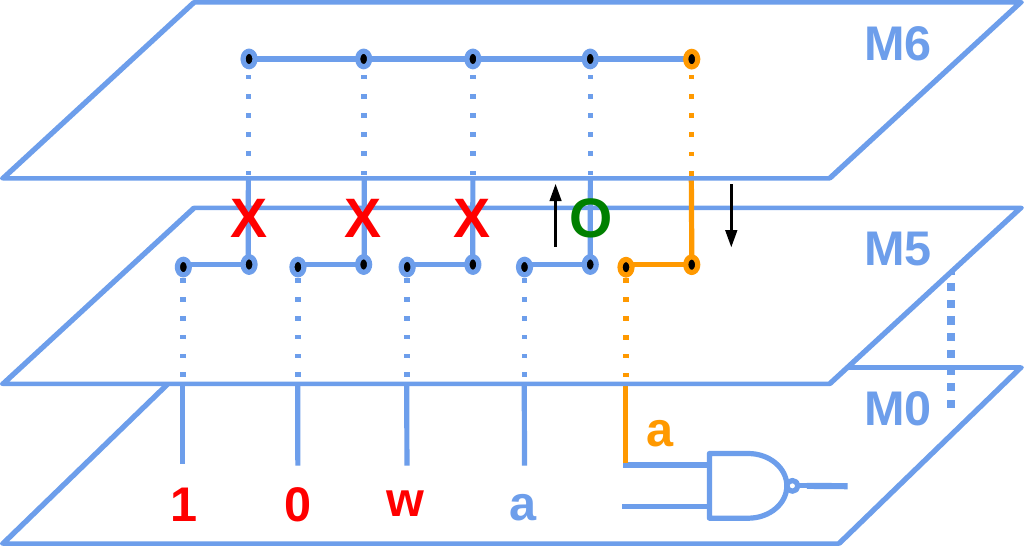}}\qquad
\subfloat[]{\includegraphics[height=3.2cm]{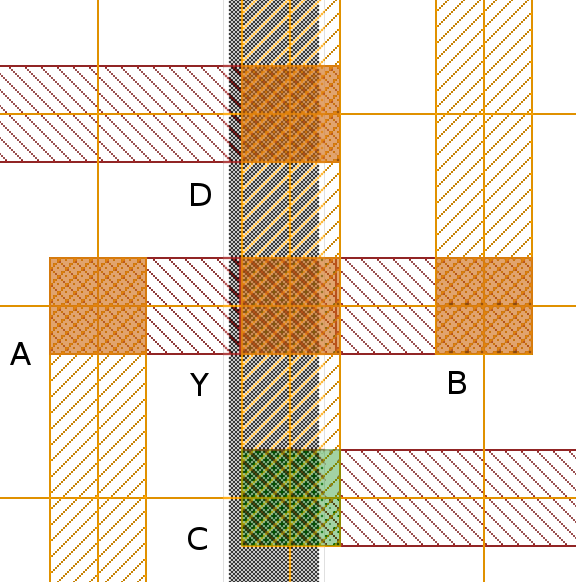}}
\smallerspacecaption
\caption{Wiring and vias for our obfuscating cell; the concept for the BEOL is illustrated in (a) and the physical-design view in (b).
		Note that the actual wiring in M5/M6 depends on which vias are dummy and which are real.
		In (a), the dummy/real vias are indicated as red crosses/green circle.
		In this example, the real net
		is labeled a, and it connects to pin C in (b); the camouflaged gate's input in (a) is wired with pin Y in (b).
		As for (b), the pins A, B, and Y reside in M6 (orange, vertical wiring), whereas pins C and D are set up in M5 (red, horizontal wiring).
		The vias (all between M5 and M6) are represented by the pins and either colored in orange (dummy vias) or in green (real via).
		Also note the cell's grey
		outline beneath the pins;
the latter
are partially located outside
the cell, which is feasible/supported.
		The minimal width of the cell is solely to ease its visual differentiation from standard cells at design time.
}\label{fig:final_primitive_layout}
\end{figure}

Given an HDL netlist, we initially synthesize, place, and route the design.
On this {
		original layout}, we then apply our transformations outlined above.
Next, we insert and wire \emph{obfuscating cells} for all the inputs of each gate to be camouflaged.
It is essential to understand that these custom cells do not impact the FEOL layer---their sole purpose is to
enable the routing of all dummy and real wires
(Fig.~\ref{fig:final_primitive_layout}).
Hence, the physical design of these custom cells is tailored for routability, while their arrangement remains
flexible and unconfined regarding the already placed standard cells (see also Sec.~\ref{sec:obfuscating_cell}).

For our final primitive,
	recall that
		there are four obfuscated wires for each input (i.e., for each obfuscating cell, Fig.~\ref{fig:final_primitive_layout}):
two wires connect to 
\emph{1} and \emph{0} (either randomly with regular TIE cells or with gates acting as TIE cells in disguise),
one wire connects with the real net, and one wire with a \emph{dummy net}.
	In case
	\emph{1} or \emph{0} is the desired input---i.e., when the gate shall implement INV, BUF, or a TIE cell in disguise---the
		real net is either replaced 
	with another unique dummy net
	or ``driven'' by another of the disguised TIE cells.

As indicated in Sec.~\ref{sec:flavors},
	we have to choose dummy nets carefully such that they are unique with respect to each gate to camouflage.
We do so by applying a local spatial search around each gate's inputs; nearby nets/wires are preferably selected to limit the routing congestion.
We also check for combinatorial loops which may have resulted during that process, and
re-select dummy nets as required.

After embedding and connecting the obfuscating cells,
we perform an \emph{ECO optimization} and legalization; the latter is also based on custom constraint rules (Sec.~\ref{sec:obfuscating_cell}).
At the same time, we re-route the design---now with all the dummy wires along with the real wires.
We perform final design closure,
   remove the obfuscating cells from the design,
   re-extract the \emph{RC} data, and finally gather PPA numbers.

\subsection{Physical Design of the Obfuscating Cell}
\label{sec:obfuscating_cell}

We implemented the obfuscating cell as a custom cell and extended the LIB/LEF files.
Here we elaborate on the cell's physical design.
\smallspaceenum
\begin{compactenum}

\item
The cell has four input pins and one output pin (Fig.~\ref{fig:final_primitive_layout}(b)).
The pins
have been set up in two metal layers: pins A, B, and Y reside
in M6,
whereas pins C and D are set up in M5.
We have chosen two different layers to
minimize the routing congestion; in exploratory experiments with all pins in M6, we observed overly high congestion, especially for LC beyond 50\%.
Note that one can easily tailor the pins for different layers as well (e.g., M7/M8), if considered useful for particular designs.
\item
The dimensions of the pins ($0.14\times0.14\mu m$) and their offsets
are chosen such that the pins can be placed directly on the respective metal layer's tracks (thin yellow grid in Fig.~\ref{fig:final_primitive_layout}(b)); this is to further minimize
		the routing congestion.
\item
We define custom constraint rules
which prevent the pins of different obfuscating cells to overlap during legalization.
These obfuscating cells
can, however, freely overlap with any standard cell without inducing routing conflicts. That is because standard cells have their pins exclusively in the lower metal layers.
\item
We leverage the timing and power characteristics of BUFX2, i.e., a buffer with driving strength of 2.  Note that a detailed library characterization is not required as the obfuscating cell
only implements BEOL wires and vias.
However, we have to set up an annotation regarding the pin capacitances
(Fig.~\ref{fig:obfus_pins_cap}).
This is essential to enable proper ECO optimization and to 
evaluate the final PPA numbers.

\begin{figure}[tb]
\centering
\includegraphics[width=.455\textwidth]{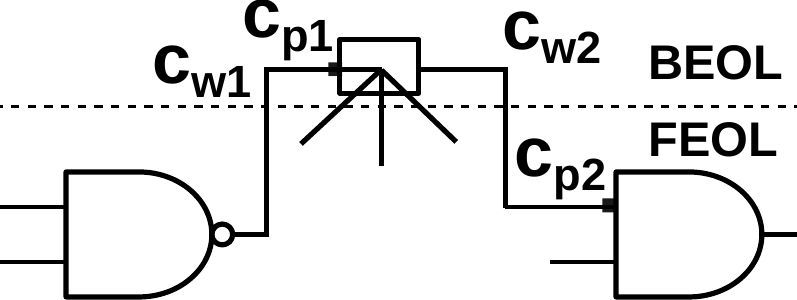}
\smallerspacecaption
\caption{The capacitance $c_{p1}$ for any input pin of the obfuscating cell
		is to be
		annotated, to account for both the wire capacitance $c_{w2}$ and the camouflaged gate's input-pin
		capacitance $c_{p2}$.
		Otherwise, the respective driver's load would be underestimated during ECO optimization.
}\label{fig:obfus_pins_cap}
\end{figure}

\end{compactenum}

\section{Experimental Investigation}
\label{sec:experiments}

\textbf{Setup for layout evaluation:} We implement our methodology as custom scripts for \emph{Cadence Innovus 15.1};
all our procedures incur negligible runtime cost.
We employ the public \emph{NanGate 45nm Open Cell
Library}~\cite{nangate11} with ten metal layers.
The PPA analysis is carried out for
0.95V, 125$^\circ$C, and the slow process corner, along with a default input switching activity of 0.2---note
that this is a rather conservative setup.
Power and timing results are obtained by \emph{Innovus} as well.
We configure the initial utilization rates (i.e., for the original layouts) such that the routing congestion remains below 1\%.

Recall that we seek to safeguard our camouflaged layouts also from
fab
adversaries and that we employ split manufacturing towards this end (Sec.~\ref{sec:concept}).
We suggest splitting at  M5,
    i.e., just beneath the Mg/MgO vias which represent our secret assets to be protected.\footnote{
Xiao \emph{et al.}~\cite{xiao15} noted that splitting at higher metal layers imposes relatively low efforts; higher layers have rather large
	pitches, which are easy and cheap to manufacture by the trusted (low-end) BEOL facility.
Besides cost and practicability, however, we acknowledge that
Wang \emph{et al.}~\cite{wang16_sm} argued that splitting at M5
	may not be secure, based on their advanced \emph{proximity attack}.
Here it is important to note that our approach
allows splitting beneath M5 without any restriction.
Moreover, we observe in exploratory experiments that our LC scheme
inherently helps mitigating such proximity attacks.
More specifically, after fully camouflaging 14 selected designs, splitting their layouts at M5, and running the attack of~\cite{wang16_sm} against the FEOL layouts,
	we observe \emph{correct connections} of only 22.5\% on average.
We believe that is because all camouflaged gates are routed
through M6 and above. Especially for large-scale LC, thus, our scheme induces a plethora of open nets (i.e., nets that are cut across FEOL and
		BEOL) which is challenging for any such proximity attack.
A more detailed study
		will be the scope of our future work.}

\textbf{Setup for security evaluation:} We implement all LC techniques proposed in~\cite{rajendran13_camouflage, wang16_MUX, zhang16, malik15} for the sake of comparison with our work.
For a fair evaluation, the same sets of gates are camouflaged across all LC techniques:
for a given benchmark, gates are randomly selected once and then memorized.\footnote{
		Besides random selection, any other technique such as \emph{maximum clique}~\cite{rajendran13_camouflage} can be applied as well.
			Somewhat surprisingly, Massad \emph{et al.}~\cite{massad15} observe that random selection is on average almost as effective.}
Ten different sets are generated for each benchmark,
ranging from 10\% to 100\% LC, in steps of 10\%.

Recall that we evaluate the
LC primitives against powerful SAT attacks~\cite{subramanyan15} which are publicly available~\cite{code_pramod}.
Note that the tool~\cite{code_pramod} was developed for logic locking but is still applicable for our study; logic locking and LC are closely
related and can be transformed into one another~\cite{yu17}.
All the
SAT attacks
are executed on a server with five compute nodes, where each node has two 14-core Intel Broadwell processors, running at
2.4 GHz with 128 GB RAM.
The CPU time-out
		(``t-o'') is set to 48 hours.

      We attribute both the runtime
	      and the \emph{growth trend for clauses}
	      as primary indicators for a design's resilience.
	     The latter is helpful for large-scale LC when monitoring the attack runtimes becomes prohibitive.
	     Specifically, the trend of clauses
indicates whether the SAT solver can simplify the structures in the camouflaged design at all~\cite{yu17}.
In case the number of clauses is continuously (and linearly) increasing,
i.e., the saturation of clauses is not reached,
we can conclude that the attack will not finalize in foreseeable time.

\textbf{Benchmarks:} We conduct extensive experiments on traditional benchmarks suites (\emph{ISCAS-85}, \emph{MCNC}, and \emph{ITC-99}) and, for the first time, also on the
large-scale \emph{EPFL MIG suite}~\cite{EPFL15}. 
For the latter, we compile the original circuits, not the MIG versions.
Selected benchmarks are reviewed in Table \ref{tab:benchmarks}; we consider 34 benchmarks in total, however.
Note that all benchmarks are
combinatorial, but our
approach can be directly applied to sequential designs as well.

\begin{table}[tb]
\centering
\scriptsize
\caption{Characteristics of Selected Benchmarks
 (Those in Italics are from the EPFL Suite~\cite{EPFL15}, Others are from Traditional Suites)
}\label{tab:benchmarks}
\smallerspacecaption
\begin{tabular}{|c|c|c|c|}
\hline
 \textbf{Benchmark
 } & \textbf{Inputs} & \textbf{Outputs} & \textbf{Gate Count
 } \\ \hline \hline
\emph{aes\_{core}}  & 789 & 668 & 39,014 \\ \hline
b14 & 277 & 299 & 11,028 \\ \hline
b15 & 485 & 519 & 10,354 \\ \hline
b17 & 1,452 & 1,512 & 36,770 \\ \hline
b22 & 767 & 757 & 33,110 \\ \hline
c7552 & 207 & 108 & 4,045 \\ \hline
des & 256 & 245 & 6,473 \\ \hline
\emph{diffeq1} & 354 & 289 & 30,584 \\ \hline
\emph{square} & 64 & 127 & 28,148 \\ \hline
\end{tabular}
\end{table}

\subsection{Layout Evaluation}
\label{sec:layout_evaluation}

\begin{figure}[tb]
\centering
\includegraphics[width=0.9115\textwidth]{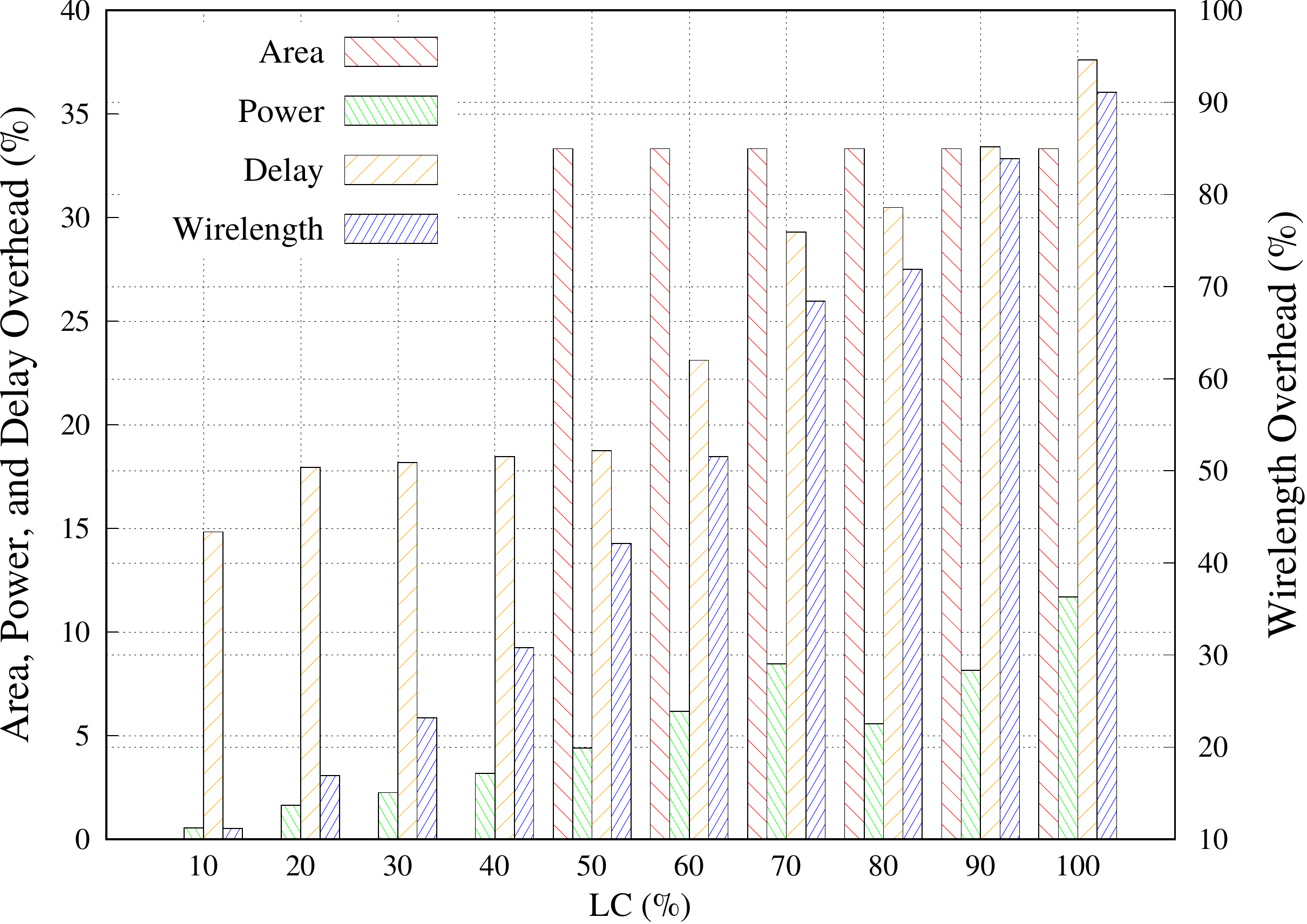}
\smallerspacecaption
\caption{Layout cost for \emph{aes\_{core}}~\cite{EPFL15} using our final LC primitive, with the
	baseline being the original layout.
		The discrete and monotonous area cost is due to a step-wise up-scaling of die outlines as needed; see also below.
}\label{fig:PPA_aescore}
\end{figure}

Here we report on general trends for PPA cost regarding our final LC primitive, and
	we also compare to previous works.
Figure~\ref{fig:PPA_aescore} illustrates the cost for
	benchmark \emph{aes\_{core}}~\cite{EPFL15}, and Table~\ref{tab:ppacomparison} reports the cost for this and other benchmarks used in this work.

\begin{table*}[tb]
\centering
\scriptsize
\setlength{\tabcolsep}{1.1mm}
\caption{Our
		GDSII-Level Cost
		in \% for Large-Scale LC (in \% of All Gates) on Selected benchmarks\label{tab:ppacomparison}
}
\smallerspacecaption
\begin{tabular}{|*{17}{c|}}
\hline
\multirow{2}{*}{\textbf{Benchmark}}
& \textbf{Utilization for}
& \multicolumn{3}{|c|}{\textbf{20\% LC}} & \multicolumn{3}{|c|}{\textbf{40\% LC}} & \multicolumn{3}{|c|}{\textbf{60\% LC}} &
\multicolumn{3}{|c|}{\textbf{80\% LC}} & \multicolumn{3}{|c|}{\textbf{100\% LC}}  \\
\cline{3-17}
& \textbf{Original Layout} & \textbf{Area} & \textbf{Power} & \textbf{Delay} & \textbf{Area} & \textbf{Power} & \textbf{Delay} & \textbf{Area} & \textbf{Power} & \textbf{Delay} & \textbf{Area} & \textbf{Power} & \textbf{Delay}  & \textbf{Area} & \textbf{Power} & \textbf{Delay}  \\ \hline
\hline
\emph{aes\_{core}} & 0.4 & 0 & 1.6 & 17.9 & 0 & 3.1 & 18.4 & 33.3 & ~~5.5 & 23.1 & 33.3 & ~~6.1 & 30.4 & 33.3 & 11.6 & 37.6  \\ \hline
b14 & 0.5 & 0 & 2.3 & ~~8.1 & 0 & 5.3 & ~~9.2 & 0 & 11.4 & 17.9 & 25.0 & 14.6 & 21.1 & 25.0 & 15.1 & 22.1  \\ \hline
b15 & 0.5 & 0 & 2.3 & 13.8 & 0 & 6.1 & 21.8 & 25.0 & ~~7.2 & 25.1 & 25.0 & ~~8.5 & 27.4 & 25.0 & 12.7 & 41.5 \\ \hline
b17 & 0.5 & 0 & 2.8 & 15.5 & 25.0 & 2.7 & 28.6 & 25.0 & ~~4.9 & 31.2 & 66.7 & ~~6.6 & 30.3 & 66.7 & ~~8.7 & 36.9 \\ \hline
b22 & 0.6 & 0 & 3.7 & ~~9.3 & 20.0 & 4.6 & 16.4 & 20.0 & ~~6.1 & 17.4 & 50.0 & 11.2 & 26.2 & 50.0 & 19.0 & 31.9   \\ \hline
\emph{diffeq1} & 0.5 & 0 & 3.5 & 14.9 & 25.0 & 8.6 & 25.8 & 25.0 & ~~6.0 & 12.3 & 66.7 & ~~8.2 & 13.6 & 66.7 & 10.2 & 16.3    \\ \hline
\emph{square} & 0.5 & 0 & 1.8 & 10.1 & 25.0 & 3.6 & 13.9 & 25.0 & ~~4.6 & 15.1 & 66.7 & ~~5.4 & 16.4 & 66.7 & ~~8.2 & 24.8    \\ \hline
\hline
\textbf{Average} & 0.5 & 0 & 2.6 & 12.8 & 13.6 & 4.9 & 19.2 & 21.9 & ~~6.5 & 20.3 & 47.6 & ~~8.7 & 23.6 & 47.6 & 12.2 & 30.2  \\ \hline
\end{tabular}
\end{table*}

\textbf{On die area:} Recall that our obfuscating cells
do not tangent the standard-cell area; the reported cost is thus concerning the \emph{die outline}.
Especially for large-scale camouflaging, we have to scale up the die outlines to mitigate routing/DRC errors.
For the sake of simplicity,
    we scale up the outlines in view of 
    decreasing the utilization rate in steps of 0.1. For
example, while relaxing the utilization from 0.5 to 0.4, the die area has to be increased by 25\%.
As this technique can be rather profuse, we like to note that a more conservative up-scaling may enable less area overhead.

The overheads are typically not more than 25\% for up to 60\% LC, whereas for 100\% LC, we note an average area cost close to 50\%.
Again, these overheads enable DRC-clean layouts even for full-chip camouflaging---we
    believe that this is a justifiable achievement.

\begin{figure}[tb]
\captionsetup[subfigure]{labelformat=empty}
\centering
\subfloat[]{\includegraphics[width=.495\textwidth]{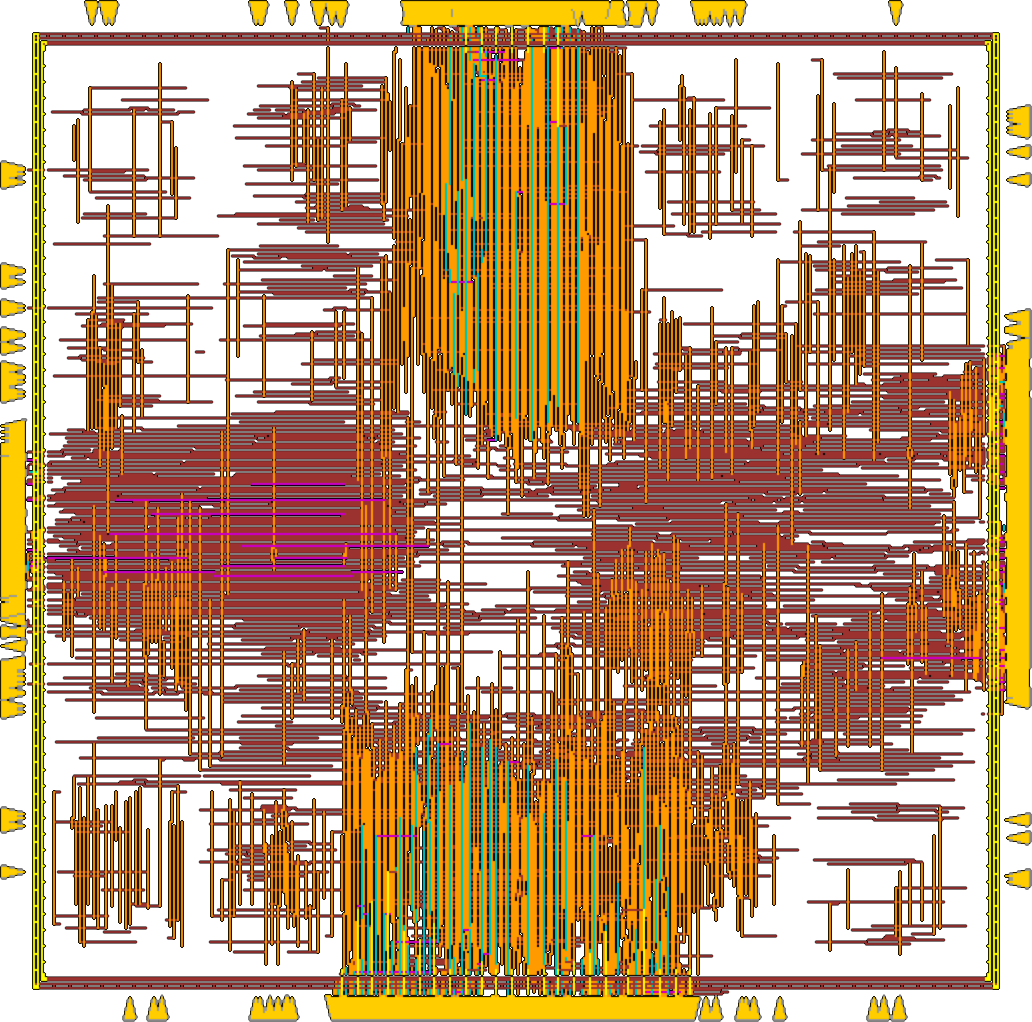}}\hfill
\subfloat[]{\includegraphics[width=.495\textwidth]{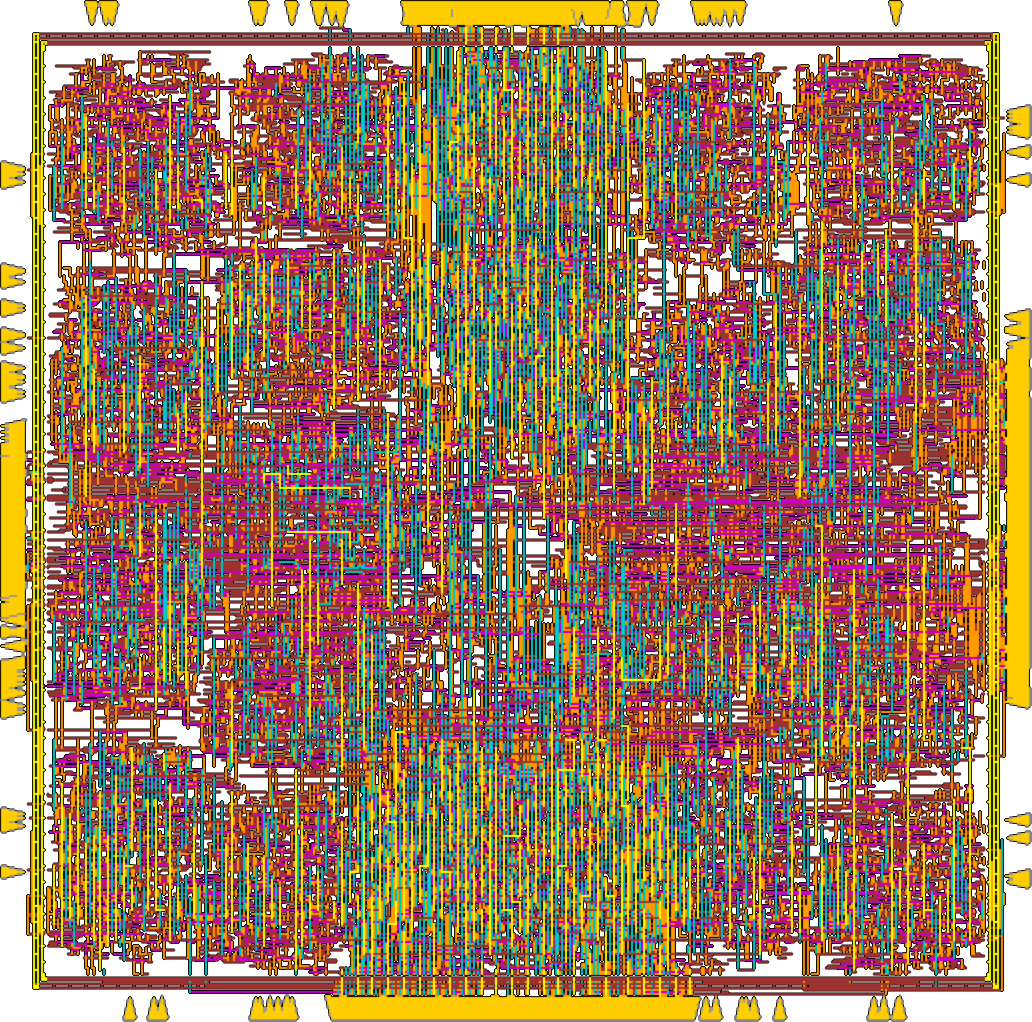}}\\
\smallerspacecaption
\smallerspacecaption
	\caption{Metal layers M5 to M10 for benchmark \emph{aes\_{core}}~\cite{EPFL15}; the original layout is on the left and the layout with 100\% LC
			is on the right.
    \label{fig:metal-AES}
	}
\end{figure}

\textbf{On power and performance:} Recall that our primitive has all its wires routed in higher metal layers; see also
Fig.~\ref{fig:metal-AES}.
Hence, an impact on power and performance is expected.

Interestingly, for camouflaging up until 60\% of the layout, the average overheads are relatively small, typically in the range of 2--7\% for power and 12--21\% for delay. The
overheads increase for larger and full-scale LC, but still follow a linear growth.
\noindent
We believe that these two trends are due to the following:
\smallspaceenum
\begin{compactenum}
\item Besides the transformed INV/BUF gates, all gates remain as is; we experience no inherent overheads for the majority of gates.
\item The relatively low resistance of the higher metal layers used by our primitive helps to limit the delay cost for LC.
\item The ubiquitous nets for the fixed values \emph{0} and \emph{1} are not switching and, thus, they neither increase power nor delay.

\item For LC beyond 60\%, the positive effects above are offset by the steady increase of the wiring for camouflaged gates, thereby raising the routing congestion.
Since routing congestion
can only be managed by re-routing in some detours, this lengthens parts of the wires further. In turn, this also impacts power and delay.

\end{compactenum}
\smallspaceenum

\textbf{Comparison on layout level:} Recall that our work is one of the very few to evaluate LC on placed and routed GDSII designs.
When contrasting to a previous study, conducted by Malik \emph{et al.}~\cite{malik15}, we observe significantly lower overheads;
such a qualitative comparison is fair as the authors use the same \emph{NanGate} library~\cite{nangate11}.
Specifically, Malik \emph{et al.\ }reported
overheads of 7.09$\times$, 6.45$\times$, and 3.12$\times$ for area, power, and delay, respectively.
It is also important to note that Malik \emph{et al.\ }implement and evaluate their approach
for one \emph{AES S-box},
which has a far lower number of gates---namely only 421---when compared to all the benchmarks we consider.
The authors indicate themselves that the
 cost will increase for larger circuits~\cite{malik15}.

For Zhang's work~\cite{zhang16}, as its MUX-based primitive is not publicly available, we implement it ourselves, and perform a detailed layout-level evaluation. For example, we observe 464\%, 638\%, and 63\% increase in
area, power, and delay, respectively, when camouflaging all the gates for the \emph{ISCAS-85} benchmark \emph{c7552}.
These numbers are 8.3$\times$, 55.4$\times$, and 2.18$\times$ higher than ours for the same scenario.

Besides advocating a provably secure LC scheme,
Li \emph{et al.}~\cite{li16_camouflaging} also propose two different LC primitives, called \emph{STF-type} and \emph{XOR-type}.
As the authors report only gate-level cost, we conduct a layout-level evaluation ourselves as well (Table~\ref{tab:ppacomparison_new}).
Here we map the numbers reported for their primitives to regular gates of the respective type---doing so is fair
	and conservative as it implies only a linear scaling.

\begin{table}[tb]
\centering
\scriptsize
\setlength{\tabcolsep}{0.7mm}
\renewcommand{\arraystretch}{1.08}
\caption{Comparison of Area (A), Power (P), and Delay (D)
		for 100\% LC on Selected Benchmarks (N/A Means Not Available)\label{tab:ppacomparison_new}
}
\smallerspacecaption
\begin{tabular}{|*{13}{c|}}
\hline
\multirow{2}{*}{\textbf{Benchmark}}
& \multicolumn{3}{|c|}{\textbf{XOR-type~\cite{li16_camouflaging}}} & \multicolumn{3}{|c|}{\textbf{STF-type~\cite{li16_camouflaging}}} &
	\multicolumn{3}{|c|}{\textbf{\cite{erbagci16}}} & \multicolumn{3}{|c|}{\textbf{Ours}}
  \\
\cline{2-13}
 & \textbf{A} & \textbf{P} & \textbf{D} & \textbf{A} & \textbf{P} & \textbf{D} & \textbf{A} & \textbf{P} & \textbf{D}  & \textbf{A} & \textbf{P} & \textbf{D}   \\ \hline
\hline

apex4 & 59.7 & ~~5.9 & 45.1 & 51.3 & 18.8 & 45.3 & N/A  & N/A  & N/A  & 50.0 & 17.5 & 41.8   \\ \hline
c432 & 51.1 & 19.3 & ~~9.8 & 38.9 & 32.6 & 24.3 & 140.0  & ~~8.0 & 96.0 & 62.5 & 25.3 & 26.8   \\ \hline
c5315 & 96.3 & 27.4 & 51.1 & 70.1 & 44.3 & 59.8 & 200.0  & 10.0 & 76.0 & 55.6 & 17.5 & 42.2   \\ \hline
c7552 & 96.6 & 30.0 & 66.8 & 59.7 & 40.0 & 63.8 & 175.0  & ~~9.0 & 90.0 & 55.6 & 11.5 & 28.9   \\ \hline
b14 & 74.8 & 14.2 & 39.6 & 57.6 & 36.6 & 39.2 & N/A  & N/A  & N/A & 25.0 & 15.1 & 22.1   \\ \hline
b17 & 71.3 & ~~3.3 & 53.0 & 53.4 & 20.9 & 58.9 & N/A  & N/A  & N/A  & 66.7 & ~~8.7 & 36.9   \\ \hline
b20 & 78.5 & 16.4 & 46.0 & 62.6 & 40.9 & 72.2 & N/A  & N/A  & N/A & 50.0 & 15.1 & 30.1   \\ \hline
\hline
\textbf{Average} & 75.5 & 16.6 & 44.5 & 56.2 & 33.4 & 51.9 & 171.7 & ~~9.0 & 87.3 & 52.2 & 15.8 & 32.7 \\ \hline
\end{tabular}
\end{table}

\textbf{Comparison on gate level:} Previous works
typically
report their cost only for small-scale LC, which arguably is their sole scope.
For
example, the MUX-based approach of~\cite{wang16_MUX} exhibits on average 50\% delay overhead and 15\% area overhead already for 5\% LC.

We also like to note that most prior art report numbers based on RTL simulations, which seems too optimistic, especially for large-scale LC.
Conservatively assuming that these numbers would still scale only linearly, the primitive
of~\cite{rajendran13_camouflage}
would incur
$\approx$600\%, $\approx$400\%, and $\approx$300\% cost on area, power, and delay, i.e., with 50\% of the gates camouflaged.
For the same scenario,
	our technique incurs only 16.7\%, 6.9\%, and 12.8\% overheads for area, power, and delay, respectively.
That is,
we attain 36$\times$, 58$\times$, and 23$\times$ lower overheads.

\textbf{Comparison with threshold-dependent LC:} Nirmala \emph{et al.}~\cite{nirmala16} recently proposed a promising concept of threshold-dependent LC switches.
Also here we apply a thorough layout-level evaluation, across 34 benchmarks, based on their numbers reported for their primitives. As a result, we
find that
this approach will incur significant layout cost for full-chip LC: $\approx$1,360\%, $\approx$1,266\%, and $\approx$100\% for area, power, and delay,
respectively. For~\cite{collantes16}, we observe a linear trend in the reported layout cost; extrapolating those numbers for full-chip LC would translate to
$\approx$78\% and $\approx$147\% for power and delay, respectively.
Finally, for
the work of Erbagci \emph{et al.}~\cite{erbagci16}, we report their numbers in Table~\ref{tab:ppacomparison_new} for comparison.

\textbf{Comparison with provably secure LC:} Recall that~\cite{xie16_SAT,yasin16_CamoPerturb,li16_camouflaging}
	rely on additional circuitry to
	protect individual, selected gates/wires.
		Such circuitry can incur a high cost, especially for area.
		For example
	in Xie \emph{et al.}'s work~\cite{xie16_SAT},
when protecting one gate/wire of the 
\emph{c7552} benchmark using 64 key bits,
the die-level area overhead we observe in our layout-level implementation of their work is close to 106\%.\footnote{
	Since these studies~\cite{yasin16_CamoPerturb,xie16_SAT,li16_camouflaging} are all based on the same principle (inserting combinatorial trees), we can also expect
		similar trends for the other schemes.}
If we were to protect more gates/wires,
the overhead would scale up accordingly.
	In contrast, when camouflaging \emph{all gates} of the same circuit, our approach incurs only $\approx$56\% overhead for the die area.

\subsection{Security Evaluation}
\label{sec:security_evaluation}

Recall that our primary objective is large-scale LC; an important observation is that this achieves \emph{practically secure LC}.
That is, by camouflaging up to 100\% of the layout, we seek to induce the highest computational effort possible for SAT attacks, at least without inserting additional \emph{provably
	secure} structures.

\textbf{On the notion of practically secure LC:}
It is not straightforward to prove beforehand to what extent large-scale LC will render a layout (practically) secure without leveraging a SAT solver's capabilities for de-camouflaging.
Li \emph{et al.}~\cite{li16_camouflaging} have shown that the effort for de-camouflaging scales \emph{on average} with both ($i$)~the solution space $C$ concerning all the possible
functionalities for the whole, camouflaged design and ($ii$)~the Hamming distance among those different functionalities.
In turn, the key reason why a further theoretical evaluation
of large-scale LC is so difficult is
that the space of $C$ (and thereby the Hamming distance as well) depends
on ($a$)~the types and count of possible functionalities for all employed LC primitives, ($b$)~the count and selection of gates to camouflage, and ($c$)~the connectivity
among all gates in the design, all at the same time.
On the one hand, for example, designs containing XOR/XNOR and/or multipliers are harder to de-camouflage in practice~\cite{yu17,subramanyan15}.
On the other hand, as with classical logic minimization, the solution space of some interconnected camouflaged gates (but not for [N]AND-trees) may be largely simplified during the SAT
search.

In short, while we can easily estimate the upper bound of $C$, it may be far from
the actual number
of functions required to consider, depending on the design.
This implies that without conducting SAT attacks (or similar techniques), there is no basis for a fair
evaluation of different LC schemes.
Hence we resort to such an empirical but comprehensive study, also to contrast our work to prior art.

As we observe in this study,
	we can indeed expect prohibitive runtimes for large-scale LC. More specifically, e.g., we extrapolate that
de-camouflaging \emph{aes\_core} (when all 39,014 gates are camouflaged using our primitive) can take approximately 108 years.\footnote{For a meaningful regression, we sample the average attack
	runtimes across 70
	different sets of randomly camouflaged gates.
}
Similar observations have also been
made by Yu \emph{et al.}~\cite{yu17}, albeit for a different primitive and the much smaller benchmark \emph{c432} (209 two-input
		gates); even this layout could not be resolved within three days.

\textbf{Comparative study on the resilience of LC:}
In Table~\ref{tab:satattackcomparison}, we list the runtime for SAT attacks
	on camouflaged
designs, contrasting our LC primitive and those
of~\cite{rajendran13_camouflage, wang16_MUX, zhang16, malik15}.
Note that these prior studies did not report on any SAT attacks. Thus, we model their primitives ourselves as outlined in~\cite{subramanyan15,yu17}.
Also, recall that we camouflage the same sets of gates across all techniques for a fair comparison.

\begin{table*}[tb]
\centering
\scriptsize
\setlength{\tabcolsep}{1mm}
\caption{Runtime for Our SAT attacks (Using~\cite{code_pramod}) on Selected Designs, in Seconds
		(Time-Out t-o is 172,800 Seconds, i.e., 48 Hours)
}\label{tab:satattackcomparison}
\smallerspacecaption
\begin{tabular}{|*{21}{c|}}
\hline
\multirow{2}{*}{\textbf{Benchmark}}
& \multicolumn{5}{|c|}{\textbf{10\% LC}} & \multicolumn{5}{|c|}{\textbf{20\% LC}} & \multicolumn{5}{|c|}{\textbf{30\% LC}} &
	\multicolumn{5}{|c|}{\textbf{40/50--100\% LC}}  \\
\cline{2-21}
& \textbf{\cite{rajendran13_camouflage}} & \textbf{\cite{malik15}}
& \textbf{\cite{wang16_MUX}} & \textbf{\cite{zhang16}} & \textbf{Our} & \textbf{\cite{rajendran13_camouflage}} & \textbf{\cite{malik15}}
& \textbf{\cite{wang16_MUX}} & \textbf{\cite{zhang16}} & \textbf{Our} & \textbf{\cite{rajendran13_camouflage}} & \textbf{\cite{malik15}}
& \textbf{\cite{wang16_MUX}} & \textbf{\cite{zhang16}} & \textbf{Our} & \textbf{\cite{rajendran13_camouflage}} & \textbf{\cite{malik15}}
& \textbf{\cite{wang16_MUX}} & \textbf{\cite{zhang16}} & \textbf{Our}  \\ \hline
\hline
 
\emph{aes\_{core}} & 703 & t-o & t-o & 162 & 6,732 & 4,779 & t-o & t-o & 470 & t-o & 6,783 & t-o & t-o & 2,304 & t-o & t-o & t-o & t-o & t-o & t-o \\ \hline

\emph{ac97\_{ctrl}} & 102 & t-o & 3,271 & 37 & 1,723 & 346 & t-o & t-o & 97 & 8,678 & 4,211 & t-o & t-o & 589 & t-o & t-o & t-o & t-o & t-o & t-o \\ \hline

b15 & 423 & t-o & t-o & 65 & 583 & 5,163 & t-o & t-o & 331 & 10,012 & 5,163 & t-o & t-o & 3,306 & t-o & t-o & t-o & t-o & t-o & t-o \\ \hline

b17 & 7,894 & t-o & t-o & 1,340 & t-o & t-o & t-o & t-o & 6,041 & t-o & t-o & t-o & t-o & t-o & t-o & t-o & t-o & t-o & t-o & t-o \\ \hline

c7552 & 39 & 1,686  & 2,429 & 12 & 68 & 160 & t-o & t-o & 103 & 8,486 & 1,589 & t-o & t-o & 746 & t-o & t-o & t-o & t-o & t-o & t-o \\ \hline

\emph{diffeq1} & t-o & t-o  & t-o & t-o & t-o & t-o & t-o & t-o & t-o & t-o & t-o & t-o & t-o & t-o & t-o & t-o & t-o & t-o & t-o & t-o \\ \hline

\emph{square} & t-o & t-o  & t-o & t-o & t-o & t-o & t-o & t-o & t-o & t-o & t-o & t-o & t-o & t-o & t-o & t-o & t-o & t-o & t-o & t-o  \\ \hline

\end{tabular}
\end{table*}

It is noteworthy that none of the layouts can be de-camouflaged within 48 hours once full-chip LC is applied. In fact, all layouts remain already
resilient just beyond 40/50\% LC.
We ran further exploratory attacks for 7 days on large-scale LC using our primitive---without observing any improvement.
We can reasonably expect other prior art, such as the XOR-type/STF-type gates of~\cite{li16_camouflaging},
to perform comparably well for large-scale LC (i.e., at least on the same benchmarks). That is because the set of possible functionalities which these prior studies implement are typically within the same range.

For a more meaningful comparison, we also consider relatively small scales for LC (10\% up to 30\%). Here, the primitive advocated by Zhang~\cite{zhang16} appears as the weakest, and the one proposed by Malik \emph{et al.}~\cite{malik15} seems the most resilient.
	Our primitive is next only to that of Wang \emph{et al.}~\cite{wang16_MUX} and that of Malik \emph{et al.}~\cite{malik15}.
	Now, it is
	also important to
	recall that our proposed technique incurs significantly smaller PPA cost than these schemes~\cite{wang16_MUX,malik15}.

In short, we argue that our proposed technique is currently the only resilient yet practical solution towards large-scale LC.

\begin{figure}[tb]
\centering
\includegraphics[width=.85\textwidth]{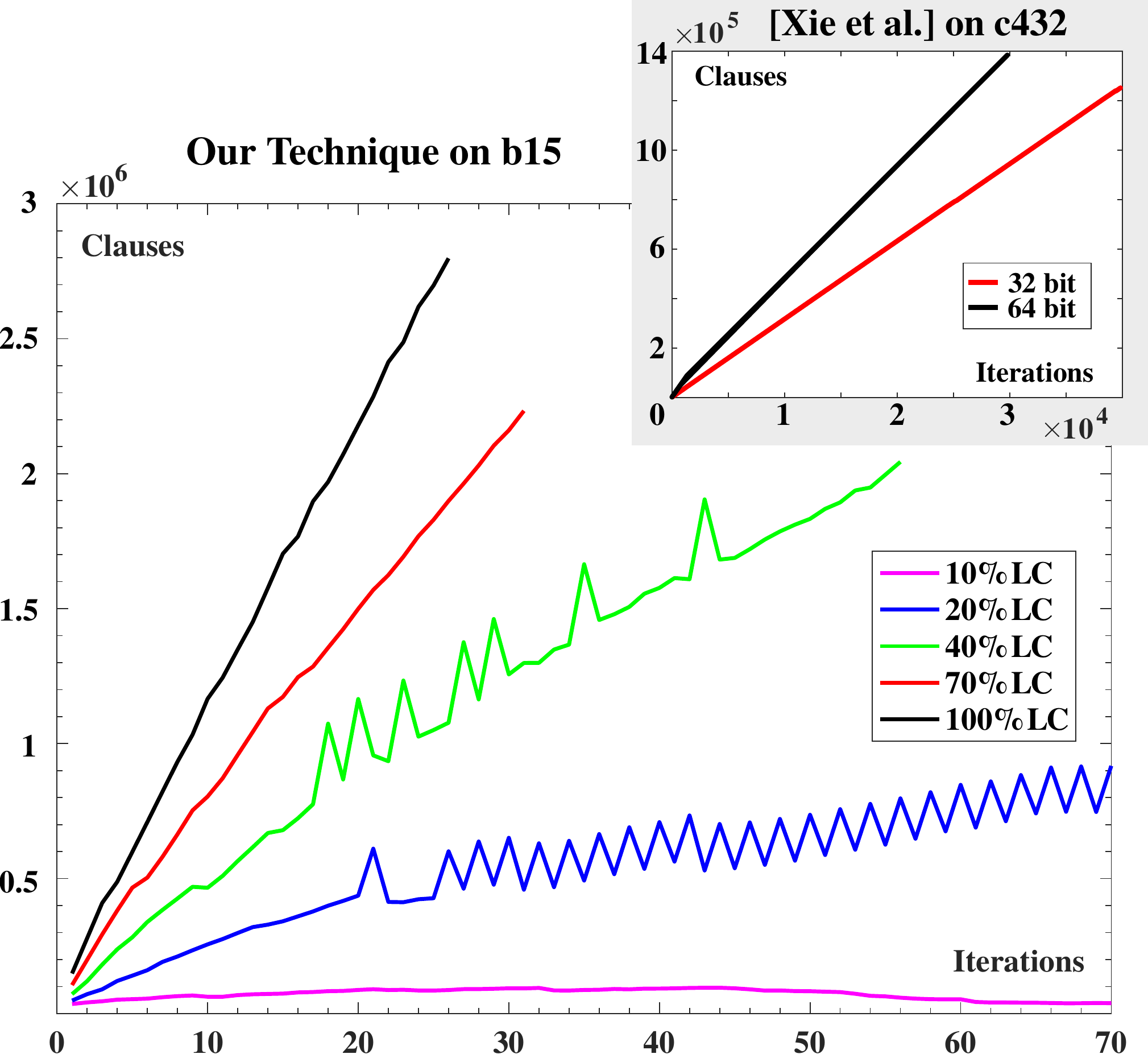}
\caption{The progress of our SAT attacks (using~\cite{code_pramod}) on
	our LC implementation (main plot) and~\cite{xie16_SAT} (inset).
		For ours, 10\% and 20\% LC is resolved;
		the larger setups incur time-out (48 hours) and excessive numbers of clauses.
			For~\cite{xie16_SAT}, neither
			setups (32- and 64-bit keys) are resolved before time-out.
}
\label{fig:SAT-progress}
\end{figure}

\textbf{On the SAT-attack effort against large-scale LC:} Recall that
	the growth of clauses hints on
any SAT attack's progress~\cite{yu17}.
When employing our primitive for small-scale LC, the number of clauses saturates at some point, and the layout can be de-camouflaged
	within 48 hours
	(Fig.~\ref{fig:SAT-progress}).
On the contrary, for large-scale LC (again, which has not been resolved),
we observe that the number of clauses increases continuously.
We found a similar trend
	when conducting attacks on~\cite{xie16_SAT} (Fig.~\ref{fig:SAT-progress}, inset).
This similarity indicates that attacks on our primitive---once it is applied at large scales---require enormous efforts, similar to attacks on SAT-hardened primitives such
as~\cite{xie16_SAT}.

Overall,
we are \emph{not} claiming that large-scale LC (using our primitive or prior art) cannot be resolved \emph{eventually} using SAT attacks.
	Rather, we provide strong empirical evidence that such practically secure LC (i.e., once 50--100\% of all gates are camouflaged) imposes a prohibitive computational cost on SAT
	solvers.
Besides, while the SAT-solver capabilities are only going to
increase, physical limits for computation will remain in any case~\cite{massad15, markov14}.

\textbf{On the resilience of~\cite{chen15}:} Since our work is inspired by~\cite{chen15} to some degree, it seems imperative to investigate its resilience as well.
We model the attack on~\cite{chen15} as outlined in Fig.~3 of~\cite{yu17}.
It can be seen from Table~\ref{tab:chen15_attack} that the SAT attacks can distinguish between real and dummy interconnects within a relatively short time, no matter how many
interconnects are obfuscated (i.e., at least concerning the delay constraints defined in~\cite{chen15}).
That is, an overly constrained
obfuscation as proposed in~\cite{chen15} cannot withstand powerful SAT attacks.
Similar observations have been reported by Yu \emph{et al.}~\cite{yu17}.

\begin{table}[tb]
\centering
\scriptsize
\caption{Runtime for SAT Attacks on~\cite{chen15}, Enabled by~\cite{code_pramod}, in Seconds
	(Columns N1--N3 Denote the Number of
	Dummy Wires Added
	While Limiting the Delay Overhead to
		0\%, 3\%, and 5\%, Respectively, as Proposed in~\cite{chen15})
}\label{tab:chen15_attack}
\smallerspacecaption
\begin{tabular}{|c|c|c|c|c|c|c|}
\hline
 \textbf{Benchmark} & \textbf{N1} & \textbf{Time} & \textbf{N2} & \textbf{Time} & \textbf{N3} & \textbf{Time} \\ \hline \hline
b14 & 30 & 7 & 36 & 9 & 55 & 11 \\ \hline
b15 & 38 & 7 & 44 & 8  & 84 & 15\\ \hline
b17 & 92 & 149 & 198  & 170  & 272 & 214\\ \hline
b18 & 265 & 2,964 & 334  & 3,223  & 518 & 3,816\\ \hline
b19 & 438 & 4,685 & 583 & 5,393  & 893 & 7,684\\ \hline
b20 & 48 & 35 & 85 & 46  & 166 & 70\\ \hline
b21 & 54 & 29 & 76 & 56  & 168 & 63\\ \hline
b22 & 76 & 58 & 113 & 79  & 191 & 128\\ \hline
\end{tabular}
\end{table}

\section{Summary and Outlook}
\label{sec:summary}

We show in this work that most techniques for layout camouflaging (LC)
		are only resilient to powerful SAT attacks once more than 50\% of the entire chip is camouflaged. We extend this call for large-scale LC 
also towards SAT-hardened and advanced LC techniques,
to thwart up-and-coming customized and/or invasive attacks.

We promote the obfuscation of interconnects as another promising avenue for LC. Towards this end, we propose and implement
BEOL-centric but generic
LC primitives (which are applicable to any FEOL node),
thoroughly contrast them to the prior art, and make our design flow publicly available in~\cite{webinterface}.
We have leveraged powerful SAT attacks and LC schemes proposed in the literature and provided a comprehensive and fair evaluation.
Further, we strive for practically relevant layout evaluation; all our camouflaged layouts are DRC-clean at the GDSII level.
 The proposed scheme is the first (to the best of our knowledge) that can deliver low-cost \emph{and} resilient full-chip
camouflaging, i.e., when considering both the layout and manufacturing
cost as well as the readily deployable protection against fab adversaries enabled by split manufacturing.

In future work, we will elaborate in more detail on the
cost and security for split manufacturing enabled by our LC primitive.

\section*{Acknowledgments}

We would like to thank
	Pramod Subramanyan (University of California, Berkeley)
	and Sergei Skorobogatov (University of Cambridge, UK)
	for their valuable inputs.
Furthermore, we are grateful to Tri Cao and Jeyavijayan (JV) Rajendran (University of Texas at Dallas) for providing their network-flow attack 
	of~\cite{wang16_sm}; this helped exploring the resilience of our primitive in the context of split manufacturing.

This work was supported in part by the Army Research Office (ARO) under Grant 65513-CS, and the
		Center for Cyber Security (CCS) at
	New York University NY/Abu Dhabi (NYU/NYUAD).
	Besides,
	this work was carried out in part on the High Performance Computing resources at New York University Abu Dhabi.

\bibliographystyle{IEEEtran}
\bibliography{main}

\end{document}